\documentclass[apj]{emulateapj}

\usepackage{natbib}
 
\submitted{Submitted to The Astrophysical Journal, August 28, 2005} 
\shorttitle{DOUBLE NEUTRON STARS: LOW ECCENTRICITIES }
\shortauthors{IHM, KALOGERA, \& BELCZYNSKI }

\begin{document}

\title{Eccentricities of Double Neutron Star Binaries}

\author{Catherine Mia Ihm and Vassiliki Kalogera}
\affil{Department of Physics \& Astronomy, Northwestern University,  2131 Tech Drive, Evanston, IL 60208}
\email{mia@alumni.northwestern.edu}

\author{Krzysztof Belczynski\altaffilmark{1}}
\affil{Department of Astronomy, New Mexico State University, Las Cruces, NM 88003}

\altaffiltext{1}{Tombaugh Fellow}

\begin{abstract}
Recent pulsar surveys have increased the number of observed double neutron stars (DNS) in our galaxy enough so that observable trends in their properties are starting to emerge.  In particular, it has been noted that the majority of DNS have eccentricities less than 0.3, which are surprisingly low for binaries that survive a supernova explosion that we believe imparts a significant kick to the neutron star.  To investigate this trend, we generate many different theoretical distributions of DNS eccentricities using Monte Carlo population synthesis methods.  We determine which eccentricity distributions are most consistent with the observed sample of DNS binaries.  In agreement with Chaurasia \& Bailes (2005), assuming all double neutron stars are equally as probable to be discovered as binary pulsars, we find that highly eccentric, coalescing DNS are less likely to be observed because of their accelerated orbital evolution due to gravitational wave emission and possible early mergers. Based on our results for coalescing DNS, we also find that models with vanishingly or moderately small kicks ($\sigma\lesssim50$\,km\,s$^{-1}$) are inconsistent with the current observed sample of such DNS. We discuss the implications of our conclusions for DNS merger rate estimates of interest to ground-based gravitational-wave interferometers. We find that, although orbital evolution due to gravitational radiation affects the eccentricity distribution of the observed sample, the associated upwards correction factor to merger rate estimates is rather small (typically 10-40\%).

\end{abstract}


\section{INTRODUCTION}
Since the discovery of PSR B1913+16, the Hulse \& Taylor binary pulsar in 1974 \citep{HT}, seven more double neutron stars (DNS) have been found.  Of all these, five systems are considered ``close" (or ``coalescing"):  PSRs B1913+16, B1534+12~\citep{B1534}, B2127+11C~\citep{B2127}, J1756--2251~\citep{J1756}, J0737--3039A~\citep{J0737A}.  Three are ``wide" (will not coalesce in a Hubble time):  PSRs J1518+4904~\citep{J1518}, J1811--1736~\citep{J1811}, J1829+2456~\citep{J1829}. A more recently discovered binary pulsar J1906+0746~\citep{J1906} is another possible close DNS, although its nature as such is not yet confirmed.
Binaries are characterized as either close or wide depending on their evolutionary timescales.  A DNS goes through an inspiral phase and eventually merges because energy is taken away from its orbit in the form of gravitational radiation.  If the time between the birth of the primordial binary and the merger of the resulting DNS is less than a Hubble time, we consider it to be ``coalescing'' (close), and wide if not.  Coalescing DNS are of interest to LIGO and other ground-based interferometers because DNS mergers are primary targets for gravitational-wave detectors.  The relativistic nature of such inspiraling DNS make them objects of extreme interest for astrophysics and relativity studies.  

Of the coalescing DNS, PSR~B2127--11C is a system with an eccentricity of $0.68$, located in the globular cluster M15 \citep{B2127}.  Star formation in the globular cluster stopped long before the birth of the recycled pulsar.  This particular binary was almost certainly created as the result of cluster dymaics. Our calculations in the present paper do not account for DNS formation in clusters, and therefore we will not include PSR~B2127--11C in our analysis. Although PSR~J1906+0746 is not yet a confirmed DNS, we include it in our analysis as such. 
\begin{table}
\centering
\caption{Observed DNS\tablenotemark{a} Binaries}
\begin{tabular}{|c|c|c|c|}
\hline
\textbf PSR & e & a ($R_{\odot}$) & $P_{orb}$ (days)\\
\hline\hline
B1913+16&   0.617   & 1.009 & 0.323 \\
\hline
B1534+12& 0.274 & 1.606 & 0.421 \\
\hline
J1756--2251&0.181&1.187&0.320\\
\hline
J0737--3039&0.088&0.610&0.102\\
\hline
J1906+0746&0.085&0.612&0.166\\
\hline\hline
  J1518+4904  &0.249&   8.634   &8.634\\
\hline
J1811--1736&0.828&14.982&18.779\\
\hline
J1829+2456&0.139&3.117&1.176\\
	\hline
\end{tabular}
\tablenotetext{a}{Eccentricity, projected semimajor axis, and orbital period of observed DNS (and recently discovered binary J1906+0746).  Information obtained from the Australia Telescope National Facility (ATNF) Pulsar Catalogue.}
\label{table:dns}
\end{table}

The orbital properties of the binaries we consider here are given in Table~\ref{table:dns}. It is evident that six of the eight observed systems have orbital eccentricities below 0.3.  The noticeable trend towards low eccentricity DNS is puzzling; one would think that a binary surviving two supernovae, believed to impart velocity kicks of up to $1000$\,km\,s$^{-1}$~\citep[e.g.,][and references therein]{Hobbs}, would typically be highly eccentric.  Part of this can be explained by circularization that occurs after the binary's first supernova.  Also, a class of low eccentricity DNS, which receive low velocity kicks at the birth of the second neutron star, has been proposed by~\citet{vdH}, as a possible explanation for the low eccentricities{~\citep[see also][]{Pods,PRPS,Dewi}}.  However, evidence for moderate to large kicks exists when when we consider the distribution of isolated pulsar velocities \citep{Hobbs, Arz, LL}. 

In this paper we examine theoretical distributions of DNS eccentricities based on various model assumptions. These distributions are generated by Monte Carlo methods using binary evolution synthesis models (\S\,2). We analyze the resulting eccentricity distributions from a number of different models (\S\,3), paying special attention to small vs.\ standard kicks and birth vs.\ present populations. We perform Bayesian statistical tests to evaluate whether any and which models are more likely, given the observed DNS sample (\S\,4). In \S\,5, we show that high-eccentricity DNS are associated with shorter merger times. Systems with short merger times are harder to detect because of their early mergers{~\citep[see also][]{GRBB2005}}. Therefore high-eccentricity DNS are not likely to contribute to the observed sample, as also concluded by~\cite{Chaur}\footnote{We note that \cite{Chaur} appeared on the LANL preprint server when we were already preparing this manuscript for submission to the Journal.}. We conclude with a discussion of our findings and their implications for DNS inspiral detections by ground-based gravitational-wave interferometers.

\section{THEORETICAL METHODS}
\subsection{StarTrack: Monte Carlo Population Synthesis}

To understand the observed sample of DNS and their eccentricities, we must provide a context for these observations by understanding the Galactic DNS population and their properties.  Theoretical models of the stellar population are generated by population synthesis methods.  Our population synthesis utilizes Monte Carlo techniques that randomly choose stellar parameters drawn from assumed distribution functions of initial conditions.  These initial parameters characterize a primordial binary, and from these initial conditions the evolution of the binary is followed, based on best current our understanding of astrophysical processes.  This is done for a large number of binaries, enough so that an adequate number of objects of interest form (in this case, DNS) for us to examine statistically.  

We use the well-tested StarTrack Population Synthesis Code to generate a large number of binaries that simulates the population and evolution of binaries in our galaxy.  A more complete description of the code can be found in~\cite{BKB} and in~\cite{Betal}.  Monte Carlo techniques are used to draw the initial parameters of the primordial binaries (masses, orbital separation, and eccentricity) from specified distributions, which are described in the next section.  Furthermore the following are chosen from appropriate distributions: (a) the time of birth for each primordial binary, chosen from 0-15 Gyr, with a constant star formation rate assumed, (b) the direction and magnitude of the kick imparted to a neutron star (NS) or black hole (BH) in an asymmetric supernova (SN) explosion, and (c) the position in the orbit where the SN takes place (relevant to eccentric orbits before the explosion).  Other aspects of the binary's evolution are based on physical laws that depend on the binary's properties with some necessary parametrizations.

We track the evolution of each binary and its physical properties at appropriate time steps from the zero-age main sequence until the formation of compact objects (white dwarfs, neutron stars, and black holes) and their subsequent orbital evolution.  Effects such as stellar winds, stable and unstable mass transfer through Roche-lobe overflow and winds, angular momentum losses, supernovae, tidal interactions of binary components, and orbital decay due to the emission of gravitational radiation are taken into account. The evolution of a binary ends when the binary merges, or when the end time of the simulation is reached (in this case, at 15 Gyr).  

\subsection{Model Assumptions}

For this paper, we look at ten different population synthesis models, denoted as A, C, E1, G1, G3, H2, L1, L2, M1, and M2 [the naming convention originates in \cite{BKB}].  Model A is our reference or {\em standard} model described next. All the other models are derived from this, with one of the synthesis model parameters varied. 

The standard model draws masses for the primordial primary (the more massive binary component) from the initial mass function $\Psi (M) \propto M^{-2.7}$ in the mass range from $4 M_{\odot}$ to $100 M_{\odot}$, the masses relevant for neutron star and black hole formation.  The binary mass ratio $q$,  defined as the ratio of the smaller mass to the larger such that $0<q\leq1$, is drawn from a flat distribution, $\Phi(q) = 1$, so that given a primary mass $M_1$, the mass of the secondary, $M_2$, is $M_2 = qM_1$.
The initial separation of a binary is chosen from a distribution that is assumed to be logarithmically flat:  $\Gamma(A) \propto \frac{1}{A}.$  
Initial eccentricities follow the thermal-equilibrium eccentricity distribution, $\Xi(e) = 2e$. For cases of dynamically unstable mass transfer and common-envelope evolution, we adopt an effective efficiency of 
$\alpha_{CE}  \times \lambda = 1$; neutron stars entering a common envelope phase are modeled taking into account hypercritical accretion. Helium star evolution is modeled in accordance to the results by~\cite{Ivanova}.  Mass transfer phases between non-degenerate stars are assumed to be non-conservative (50\% of the transfered mass is lost from the binary). This lost mass is assumed to carry a specific angular momentum equal to $2\pi jA^{2}/P$ (i.e., $j=1$). 
For supernova kicks imparted to neutron stars at birth we adopt the {\em joint} Maxwellian velocity distribution derived by \cite{Arz} with characteristic velocities $\sigma_1 = 90$ km/s (weighted at 40\%) and $\sigma_2 = 500$ km/s (weighted at 60\%). Hereafter we refer to this distribution of kicks as the ``Arzoumanian kick distribution''. 

The distinguishing characteristics of the other 9 models are described in Table~\ref{table:StartrackModels}. With this set of models we vary the values (within reasonable ranges) of a number of stellar and binary evolution parameters that are known to affect DNS formation significantly. For more information on the specific choices of values for both the standard and these models, see \cite{BKB}.

\begin{table}
	\caption{Model Assumptions}
	\centering
\begin{tabular}{ll}
	\hline \hline
\textbf Model & Description	\\
	\hline
A	&	Standard	 (see text for details) \\ & \\
C	&	No hypercritical accretion onto NS/BH \\ & in common envelopes	\\ & \\
E1	&	Effective common-envelope efficiency \\ & $\alpha_{CE}  \times \lambda = 0.1$\\ & \\
G1,G3	&	Stellar winds reduced by $f_{wind} = 0.5, 0.2$	\\ & \\
H2	&	Helium giants do not develop \\ & deep convective envelopes \\ & \\
L1,L2	&	Angular momentum of material lost in \\ 
 & non-conservative mass transfer: $j = 0.5,2.0$ \\\ & \\
M1,M2	&	Initial mass ratio distribution:	$\Phi(q) \propto q^{-2.7},q^{3}$	\\
	
	\hline
	
\end{tabular}
\label{table:StartrackModels}
\end{table}

As a possible explanation of the low eccentricities, it has been suggested that a class of neutron stars forming in DNS systems that received small-magnitudes kicks is responsible for the observed low eccentricities \citetext{see \citealp{vdH}, see also \citealp{Pods,PRPS,Dewi}}. To investigate this suggestion statistically we created another 3 sets of models which each adopt a different NS kick distribution (other than the Arzoumanian kick distribution): (i) a set of (unphysical in our opinion) models with zero NS kicks (all SN events are assumed to be symmetric); (ii) a set with a single Maxwellian of $\sigma=10$\,km\,s$^{-1}$; and (iii) a set with a single Maxwellian of $\sigma=50$\,km\,s$^{-1}$. With these choices we hope to examine the effect of zero or low kick magnitudes on DNS orbital eccentricities. Each of these three kick distributions are adopted for all ten models listed in Table~\ref{table:StartrackModels}.  

We note that the suggestion in favor of low kicks made by~\citep{vdH,Pods} is connected to NS forming from helium-rich progenitors with masses up to $\simeq 3.5$\,M$_\odot$~\citep[H-rich progenitors in the mass range $8-13$\,M$_\odot$; for details see][]{Pods}. Examination of the helium-rich DNS progenitors in all our simulations reveals that the vast majority (typically above 90\% and in some cases close to 100\%) have masses $\lesssim 3.5$\,M$_\odot$. Therefore our consideration of models with low kicks for all NS is consistent with the suggestions by~\citet{vdH} and \citet{Pods}. Furthermore the models with perfectly symmetric explosions provide a clean case that allows us to track the qualitative behavior of DNS eccentricities to pre-SN properties, as discussed in what follows.

\section{ECCENTRICITY DISTRIBUTIONS}
\begin{figure}[!t]
\epsscale{.80}
   \plotone{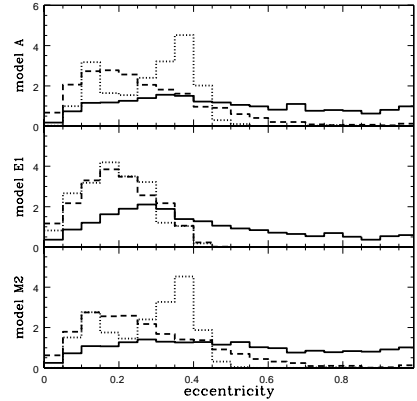} 
   \caption{Eccentricity distribution functions for coalescing binaries at birth for three of the ten models in Table~\ref{table:dns} and for three different NS kick assumptions: Arzoumanian kicks (solid lines), Maxwellian with $\sigma=50$\,km\,s$^{-1}$ (dashed lines), and zero kicks (dotted lines).  Only close binaries are considered.}
   \label{fig:ecoal}
\end{figure}
From the large populations of binaries we evolve, we consider all of the DNS systems formed and focus on the resultant DNS eccentricity distrbution functions, for each model. We typically generate $\sim30$ DNS for every $10^5$ primordial binaries evolved, though in some cases $10^5$ primordials will only produce a few relevant objects.  In short, generating reliable statistics for a variety of different models is computationally expensive and requires much time and patience. We make sure that for each model considered here we typically have model samples of $2,000-3,000$ DNS at birth and at least $\sim200$ DNS for models that have atypically low DNS formation rates (e.g., L2, M1).

\subsection{Zero, Low, and ``Standard'' Neutron-Star Kicks: Eccentricity Distributions at Birth}

\begin{figure}[!t]
\epsscale{.80}
   \plotone{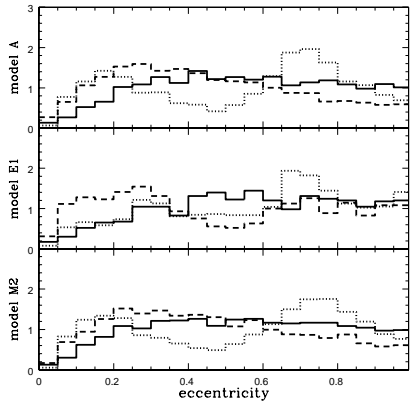} 
   \caption{Eccentricity distribution functions for wide binaries at birth. Line types as in Fig.\ 1.}
   \label{fig:enon}
\end{figure}

In Figs.~\ref{fig:ecoal} and~\ref{fig:enon} we present the eccentricity distribution functions at DNS birth for three of the models considered in our study, and for close and wide DNS, respectively. Before we proceed with our statistical analysis and comparison to the observed sample, there are a number of qualitative conclusions we can draw from these distributions. 

In the case of coalescing DNS (Fig.~\ref{fig:ecoal}), we find that the eccentricity distributions are distinctly different for the models with the Arzoumanian (``standard'') kick distribution and those with zero or low kicks. In the first case the distribution covers the full range of values up to unity, and for most models it is almost flat across the range. As kick magnitudes decrease the distributions appear to be restricted to a low range of values, most typically below $e=0.6$. This behavior appears to be consistent across all ten binary evolution models, independent of the specific model assumptions, and it is {\em qualitatively} consistent with the suggestion made by \cite{vdH} that very small NS kicks could be responsible for the low eccentricities in the observed sample. 

When supernova explosions are perfectly symmetric or have very small kicks ($\sim 10-20$\,km\,s$^{-1}$) DNS eccentricities are determined by mass loss at collapse. The qualitative behavior seen in the results of our numerical simulations can be understood based on the analytical relations that govern the orbital dynamics in this case. Following~\citet{K}, we characterize the mass loss at NS formation using the parameter $\beta$ defined as: 
 \begin{equation}
 \beta \equiv \frac{M_{NS}+M_2}{M_1+M_2},
 \end{equation}
 where $M_1$ and $M_2$ are the masses of the exploding progenitor and the companion, respectively, just prior to the explosion, and $M_{NS}$ is the mass of the resulting neutron star.
 
 Then the eccentricity after symmetric explosions assuming circular pre-SN orbits is 
 \begin{equation}
 e = \frac{1-\beta}{\beta}.
 \end{equation}
It is in the models of purely symmetric or very small kick magnitudes that the distribution of helium-star progenitors to coalescing DNS is revealed. 

The result of eccentricities being constrained to values smaller than $0.6$ clearly implies tightly constrained amounts of supernova mass loss as well, for these models: $\beta \gtrsim 0.6$, i.e., up to 40\% mass loss relative to the pre-SN total binary mass. Such a constraint is confirmed when we examine our numerical results for $\beta$. For NS masses that are all equal to a constant value (e.g., $1.4$\,M$_\odot$), this constraint would imply a very narrow range of helium-mass progenitors ($\lesssim 2.2$\,M$_\odot$). Such a tight upper limit on NS progenitor masses is marginally consistent with stellar evolution models that indicate a minimum of $2.1$\,M$_\odot$ for electron-capture collapse~\citep{Habets}. For the wider range of NS masses considered in our simulations (up to 2\,M$_\odot$), we find that helium-star progenitors up to 3-4\,M$_\odot$ become possible. This mass distribution is determined by the binary evolution prior to the supernova event leading to DNS formation; as such we find that it is not greatly affected by the presence of significant kicks or not. Their presence though affects the DNS eccentricities; the stronger kicks are relative to the pre-SN orbital velocities, the more eccentricities spread to values over the whole range from 0 to 1 and the one-to-one connection between eccentricity and SN mass loss is lost.

For wide DNS (Fig.~\ref{fig:enon}), the behavior of the eccentricity distributions is much less dependent on the presence of kicks. We find that the SN mass loss amounts cover a wider range with helium-star progenitors as massive as $8-10$\,M$_\odot$. As a result, eccentricities cover the full range up to unity with no apparent qualitative difference or bias towards low values. 

It is also interesting to note that for both close and wide binaries, in the case of zero NS kicks a double-peak structure is most clearly evident. Further investigation of this structure reveals that it is due to a double-peak structure in the iron core masses of helium stars, the NS progenitors~\citep[as first revealed and discussed in detail by][]{TWW}. This structure propagates to the eccentricity through the mass-loss during the second supernova explosion in the systems: the range of helium-star masses is narrow, and therefore the double-peak structure persists in the distribution of mass amounts lost in the explosions; since the eccentricities for symmetric explosions is directly proportional to this SN mass loss, the double-peak signature is imprinted on the eccentricity distribution too. Once kicks are introduced (even of small magnitude) this structure is ``washed-out'' due to the randomizing effect of kicks. It turns out that this behavior has no effect on our statistical analysis that follows and therefore we do not discuss this in any further detail. 

\subsection{Birth vs. Present Distributions}

\begin{figure}[!t]
\epsscale{.80}
   \plotone{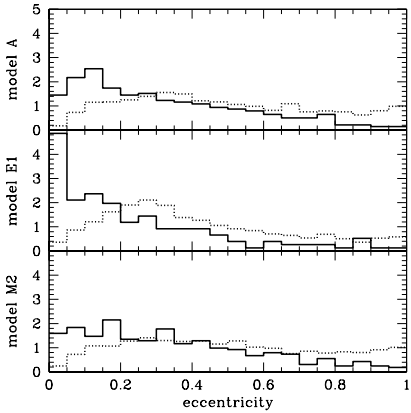} 
   \caption{Eccentricity distribution functions of coalescing binaries for three of the models considered in our study, using Arzoumanian kicks. DNS populations are shown at birth (dotted lines) and at present ($10$\,Gyr, solid lines).}
   \label{fig:eprescoal}
\end{figure}

\begin{figure}[!t]
\epsscale{.80}
  \plotone{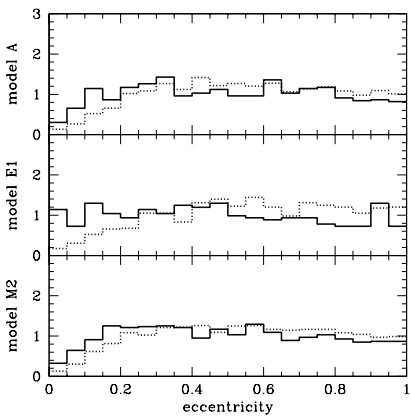} 
   \caption{Eccentricity distribution functions for wide binaries. Line types as in Fig.\ 3.}
   \label{fig:epresnon}
\end{figure}

We next compare the DNS eccentricity distributions (using models with Arzoumanian kick distributions) at two different times in their evolutionary history: at {\em birth} (the time when the second NS is formed) and at {\em present} (shown for three of our models in Fig.~\ref{fig:eprescoal},~\ref{fig:epresnon}), i.e., at $10$\,Gyr since the formation of the first primordial binaries in the simulations. Emission of gravitational radiation and resultant orbital inspiral, circularization, and mergers for the most short-lived DNS change the binaries' orbital eccentricities as time progresses. We consider the present distributions a more realistic representation of the Galactic DNS population, since binaries must have experienced this kind of orbital evolution after their birth. In our simulations, each primordial binary is assigned (randomly) a formation time, and is subsequently tracked at regular time intervals so that it is possible to know the binary properties binary at a particular time. 

It is evident that, for coalescing binaries (Fig.~\ref{fig:eprescoal}), there is a noticeable qualitative change of the present distributions towards lower eccentricities. This is in agreement with suggestions made and results obtained by \cite{Chaur}. On the other hand, for wide DNS (Fig.~\ref{fig:epresnon}) and for most binary evolution models, there is no clear evidence for significant evolution of the eccentricity distribution with time. We assert that this difference is due to the fact that the ages, i.e., time elapsed from their birth to the present, of most DNS binaries are short compared to the typical timescale relevant to orbital evolution due to gravitational radiation. Clearly this is consistent with the fact that these systems have such properties that do not coalesce within a Hubble time. 

\section{STATISTICAL ANALYSIS}

In what follows we present the implementation of a Bayesian statistical analysis, following~\cite{Finn}. This method allows us to compare the model predictions with the small-number observed samples, and calculate the likelihood of each of the models we consider here.  

\subsection{Method}

\subsubsection{Observed Eccentricities}

A priori it is not clear that the measured DNS eccentricities are indeed the true values, since measurements are associated with errors (no matter how small). Therefore, we assume that the measured eccentricities are distributed normally (following a Gaussian distribution) about the true value of each binary's eccentricity:  
 \begin{equation}
 P(e| \hat e,I) = \frac{exp \left[-\frac{1}{2} \left(\frac{e-\hat e}{\sigma_e}\right)^2\right]}{\sqrt{2\pi}\sigma_e}.
 \end{equation}  
 Here, $e$ corresponds to the observed eccentricity, $\hat e$ is the true eccentricity, and $\sigma_e$ is the reported uncertainty in the measurement.  $I$ is a parameter that encompasses all the unspecified assumptions that are also inherent in the observations.  Note that we do not really know $\hat e$, the true value; we only know $e$ and $\sigma_e$, which are the observations and errors of the eight DNS in the observed Galactic sample (Table~\ref{table:dns}). Of course, given the small errors associated with these measurements, the probability that the true value is very much different than the measured value is very small (lies in the tail of the Gaussian distribution). $P(e| \hat e,I)$ will be used to describe the probability of individual eccentricities in our Bayesian analysis, which is described in the next section.

\subsubsection{Bayesian Statistics}

In what follows we describe how we can use the observed sample and a set of theoretical eccentricity distributions to calculate the conditional probability that the observed sample is drawn from a particular theoretical model. In Bayes' law of conditional probabilities: 
 \begin{equation}
 P(M|D,I) = \frac{P(D|M,I)}{P(D,I)}P(M,I),
 \end{equation}  
 where $P(M|D,I)$ is the {\em posterior} probability, i.e., the probability associated with a model $M$, given the data set $D$;  $P(D|M,I)$ is the {\em likelihood} of the data set, given a model and we will denote by $\Lambda(D)$. $P(M,I)$ is the {\em prior} probability of the model, and $P(D|I)$ a {\em normalizing factor} for $P(M|D,I)$. This normalizing factor $P(D|I)$ is a constant so that $\int_M P(M|D,I)\,dM = 1$, and it is unknown at this point in the analysis.
 
Concerning the theoretical models considered here, we have no specific prior bias against one or another. In other words, all models $M$ are considered equally likely a priori (in the absence of data), so $P(M,I)$ must simply be a constant.  Since this prior probability is a constant, it can be absorbed into the unknown normalizing factor $P(D|I)$, so that 
 \begin{equation}
 P(M|D,I) = C \cdot \Lambda(D),
 \end{equation} 
 where $C=P(M,I)/P(D|I)$, $C$ makes $P(M|D,I)$ a properly normalized probability density, and $D$ in our case is the set of observed eccentricities, which we will denote as $e_s$.

Since the observations and property measurements for each DNS in the sample are independent, the likelihood of the complete set of measured eccentricities is equal to the product of the individual measurement likelihoods:  
 \begin{equation}
 \Lambda(e_s) = \prod_i P(e_i|M,I).
 \end{equation}
 $P(e_i|M,I)$ can be expressed as a convolution integral of $P(e_i|\hat e,I)$ over all possible $\hat e$ values, given a model $M$:  
 \begin{equation}
 P(e_i|M,I) = \int_{\hat e=0}^{\hat e=1} P(e_i|\hat e,I) \times P(\hat e|M,I)  d\hat e.
 \end{equation}  
 $P(e_i|\hat e, I)$ is just the Gaussian expression given in \S\,4.1.1 for the distribution of measured values given a ``true" value, and $P(\hat e|M,I)$ is the distribution of ``true" eccentricities, which we have generated from the Monte Carlo simulations of DNS binaries for each model.  Consequently, for each model, we have at hand all the ingredients we need to calculate the likelihood of the data set $\Lambda(e_s)$, using eqs.\ (4) and (5). The relevant integral is calculated numerically using our model distributions. Since $\Lambda(e_s)$ is strictly proportional to the posterior probability of the model $P(M|D,I)$ (see eq.\ [3]), we do not have to calculate the constant $C$, and instead we can evaluate the various models based on their likelihood values $\Lambda(e_s)$. 
The results of this evaluation is described next. 

\begin{table*}
\centering
\caption{Likelihood Values \\}
\begin{tabular}{|r||c|c|c|c|c||c|c|c|c|c|}
\hline
 &\multicolumn{5}{c||}{Close}&\multicolumn{5}{c|}{Wide}\\

\cline{2-11}
Model &
$\Lambda_{birth}\tablenotemark{a} \times 10^{43}$& 
$\Lambda_{Arz}\tablenotemark{b} \times 10^{43}$&
$\Lambda_{zero}\tablenotemark{c} \times 10^{43}$&
$\Lambda_{10}\tablenotemark{d} \times 10^{43}$&
$\Lambda_{50}\tablenotemark{e} \times 10^{43}$&
$\Lambda_{birth}\tablenotemark{a} \times 10^{22}$&
$\Lambda_{Arz}\tablenotemark{b} \times 10^{22}$&
$\Lambda_{zero}\tablenotemark{c} \times 10^{22}$&
$\Lambda_{10}\tablenotemark{d} \times 10^{22}$&
$\Lambda_{50}\tablenotemark{e} \times 10^{22}$\\
\hline\hline

A  &5.00&57.46&0.00&0.00&14.30
&1.34&2.84&3.94&6.40&3.20\\
C &4.42&32.99&0.00&0.00&87.11
&1.01&2.31&4.85&7.05&2.99\\
E1 &12.76&35.20&0.00&0.00&0.00
&0.99&2.05&1.27&3.47&9.17\\
G1 &4.11&77.71&97.75&0.00&85.93
&1.07&1.55&4.66&4.97&2.46\\
G3  &9.28&88.56&0.00&0.00&139.96
&1.28&1.09&5.83&6.42&2.96\\
H2  &6.70&34.93&0.00&0.00&0.00
&1.90&2.30&4.63&6.10&2.43\\
L1  &4.32&93.40&91.08&0.00&52.67
&5.62&2.44&4.27&5.65&3.76\\
L2  &5.64&6.16&0.00&0.00&0.00
&1.13&2.37&4.27&7.65&2.45\\
M1  &5.67&77.61&0.00&0.00&0.00
&1.42&2.53&4.27&10.68&3.81\\
M2  &5.39&52.64&0.00&0.00&58.38
&1.69&2.39&5.26&6.71&3.43\\
\hline
\end{tabular}
\tablenotetext{a}{$\Lambda$ values at birth for Arzoumanian kick distribution.}
\tablenotetext{b}{$\Lambda$ values at present for Arzoumanian kicks.}
\tablenotetext{c}{$\Lambda$ values at present for zero NS kicks.} 
\tablenotetext{d}{$\Lambda$ values at present for a $\sigma=10$\,km\,s$^{-1}$ Maxwellian kick distribution.}
\tablenotetext{e}{$\Lambda$ values at present for a $\sigma=50$\,km\,s$^{-1}$ Maxwellian kick distribution.}
\label{table:lambda}
\end{table*}

\subsection{Statistical Results} 

The full set of calculated likelihood values for various models are shown in Table~\ref{table:lambda}. We note that $\Lambda$ values are {\em not} normalized, so their absolute values have little meaning. However it is their comparative values that allow us to address the questions of interest to us: given the observed sample of measured DNS eccentricities, (i) are the birth or present populations more likely? (ii) which of the ten binary evolution models considered here more likely? (iii) are ``standard'' kick magnitudes or small or zero kicks more likely? To address each of these questions we examine the likelihood (or {\em odds}) ratios for the models relevant in each case and compare these ratios to unity. 

\subsubsection{Birth vs. Present Distributions}

\begin{figure}
	\centering
 \plotone{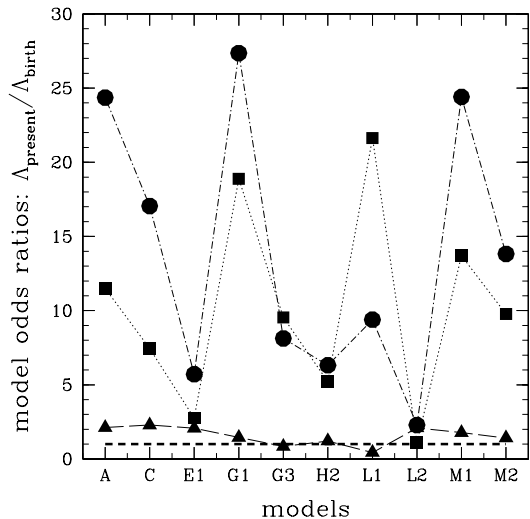} 
   \caption{Model odds ratios for the present vs. birth distributions for each of the ten binary evolution models. The squares correspond to close DNS, the triangles to wide DNS, and the circles to all systems combined.}
   \label{fig:oddspres}
\end{figure}

To quantitatively compare our results for the present and birth distributions we present the model odds ratios $\Lambda_{\rm present}/\Lambda_{\rm birth}$ as a function of the binary evolution models (Fig.~\ref{fig:oddspres}). It is evident that for the majority of the models, the ratios exceed unity often by factors of more than 2 or as high as 20. We note that for wide DNS the ratio values ``hover'' within a factor of 2 from unity indicating that there is little or insignificant evolution from birth to present for these wide DNS binaries (as already noted in \S\,3.2).  

We conclude that the models for coalescing DNS provide clear evidence that orbital evolution due to gravitational radiation is important.  When allowed to evolve, the roughly uniform eccentricity distribution at birth shifts to favor lower eccentricities at present.  Four of the five coalescing DNS inhabit the lower eccentricity part of the distributions,  so the shift towards lower eccentricities usually raises the likelihood of present model distributions over those at DNS birth. This is in full agreement with the conclusions reached by \citep{Chaur} based on a different analysis and arguments.  On the other hand the lack of strong bias in favor of the present distributions from the wide DNS models is understood, given their long orbital evolution timescales. 

Two exceptions to high odds ratios in the coalescing systems appear for models E1 and L2  (these, respectively, assume very low common-envelope efficiency and large specific angular momentum carried by mass lost from the binary in non-conservative mass transfer phases).  For model E1 (see Fig.~\ref{fig:eprescoal}), the eccentricity distribution already favors low $e$ at birth, such that when allowed to evolve, much of the DNS population possesses $e<0.05$, a region where there are no observed DNS.  We also see that the probability of seeing DNS at $0.2<e<0.3$ actually decreases, contributing to lowering the "present" likelihood.  In model L2, the odds ratio close to unity is an indication of little evolution with time for the DNS formed under L2 assumptions combined with a negligible change in the PDF in the range $0.2<e<0.3$.  

\subsubsection{Preferred Binary Evolution Models}

\begin{figure}
\centering
\plotone{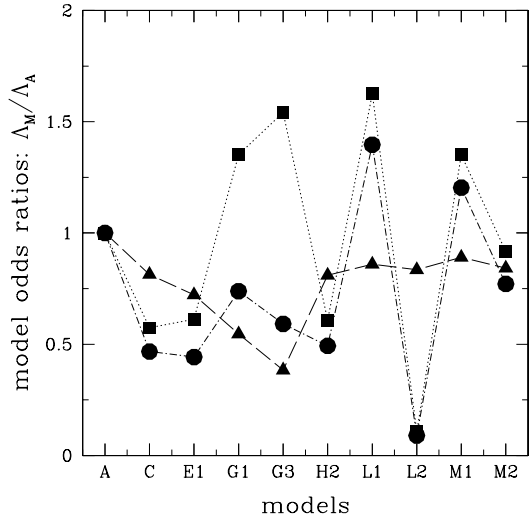}
\caption{Model odds ratios for present distributions and for all models normalized to the standard model A: $\Lambda_{\rm present, M}$ values for all models, normalized to $\Lambda_{\rm present, A}$.  Again, the squares mark close binaries, the triangles wide, and dots combine both close and wide binaries.}
\label{fig:oddsM}
\end{figure}

Guided by our conclusion that the present distributions are more likely, given the measurements, we compare the present likelihoods of the various binary evolution models to that of the standard model A (see Fig.~\ref{fig:oddsM}). 

It is interesting to note that for the majority of the models the ratios are well within a factor of 2 from unity. There is only one exception and this is model L2 with the assumed high specific angular momentum of mass lost in non-conservative mass transfer: its odds ratios for the coalescing population compared to model A is lower than 0.25. We conclude that, given the current size of the sample of measured DNS eccentricities, it is not possible to distinguish between binary evolution models; the great majority of the ones considered here appear quantitatively consistent with the eccentricity measurements, with the sole exception of L2. 

\subsubsection{Zero and Low Kick Distributions}

\begin{figure}
	\centering
   \plotone{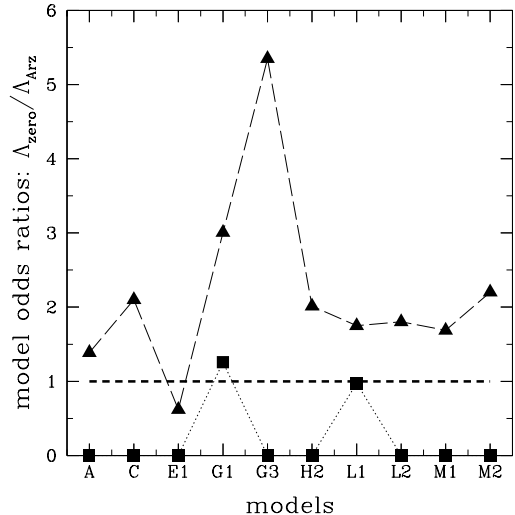} 
   \caption{Model odds ratios, $\Lambda_{zero}/\Lambda_{Arz}$ for all models.  The squares mark the close binaries; the triangles correspond to wide systems} 
   \label{fig:oddszero}
   \end{figure}

Last we examine whether the origin of the low eccentricities is related to unexpectedly low or zero kicks imparted to NS at birth. 
In Table~\ref{table:lambda} we list the likelihood values for all ten models and for 
models with zero or low kick-magnitude distributions at present. 
In Fig.~\ref{fig:oddszero} shows the model odds ratios for the zero kicks vs. the Arzoumanian kick distribution.

In the case of wide DNS the ratios for most of the models are within or very close to a factor of 2 from unity, with the exception of two models (G1 and G3). This could tentatively be used as an argument in favor of the hypothesis of very small NS kicks. 
 However, it is most interesting to examine the case of coalescing binaries: the likelihood values are {\em exactly equal to zero}, for eight models with zero kicks, for all ten models with kicks drawn from a $\sigma=10$\,km\,s$^{-1}$ Maxwellian, and for four models with kicks drawn from a $\sigma=50$\,km\,s$^{-1}$ Maxwellian. As a result all these models are rendered {\em highly unlikely}. The reason for these values being identically zero can be traced back to the properties of the Hulse-Taylor binary PSR~B1913+16: it has an eccentricity of 0.617 and all the models above have distributions such that the probability for $e\gtrsim 0.5$ is identically zero (see Fig.~\ref{fig:ecoal}). Therefore, these models cannot allow for the existence of PSR~B1913+16, and consequently their likelihood becomes equal to zero. 

Based on the results for coalescing DNS, we conclude that models with vanishingly or moderately small kicks ($\sigma\lesssim50$\,km\,s$^{-1}$) are inconsistent with the current observed sample if our assumptions about the masses of exploding stars are correct.
 
\section{CONCLUSIONS AND DISCUSSION}

We address the question of the origin of the low eccentricities measured for observed DNS binaries (close and wide) that has attracted considerable attention in the recent literature \citep{vdH,Chaur}. We use DNS population models and associated predictions for eccentricity distributions along with a Bayesian statistical analysis and quantitative comparison with the observed sample.  Assuming that all model DNS binaries can be detected as binary pulsars, we conclude that for the case of close (coalescing) DNS binaries, models with vanishingly or moderately small kicks ($\sigma\lesssim50$\,km\,s$^{-1}$) are inconsistent with the current observed sample (and specifically the Hulse-Taylor binary pulsar). We also examine the influence of orbital evolution due to gravitational radiation from the time of DNS birth to the present. We conclude that model distributions 
of DNS eccentricities {\em at present} are significantly more likely, given the observed sample. Our conclusion holds for the great majority of the models examined, unless the birth properties are such that there is very little orbital evolution from birth to the present. These results are in agreement with the conclusions obtained by \citet{Chaur} previously based on different methods of analysis. 

Here, we focus on just the distributions of eccentricities derived from simulations, since we are just interested in the origin of the low values, and not to explicitly constrain binary evolution model assumptions. In forthcoming papers we will consider the much broader questions of which binary models are most consistent with the complete set of observed DNS properties, such as orbital separations, masses, pulsar ages, in addition to eccentricities. Nevertheless we note that at a qualitative level, the range of orbital separations predicted by the {\em StarTrack} models for the DNS population is $1-6$\,R$_\odot$ for the majority of coalescing DNS, in good agreement with the measured values; distributions of such DNS orbital properties from StarTrack have been published in the past \citep{BKB}. 

\begin{figure}[!t]
  \plotone{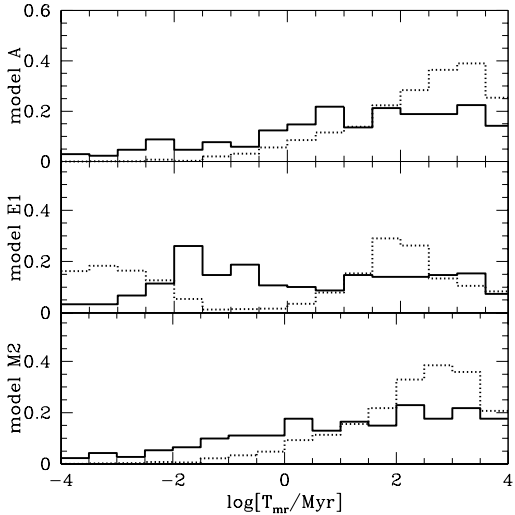} 
   \caption{The birth distributions of Tmr for models A, E1, and M2, for two eccentricity ranges: 0-0.8 (dotted lines) and 0.8-1.0 (solid lines).   Arzoumanian kicks assumed.}
   \label{fig:tmr}
\end{figure}

As also noted by \citet{Chaur}, some fraction of the systems that have high eccentricities at birth actually end up merging very soon after (the rest of the high-eccentricity systems, evolve to lower eccentricities without necessarily merging completely by the present). 
As seen it is evident in Fig.~\ref{fig:tmr}, high eccentricity systems at birth ($0.8<e<1.0$) tend to favor shorter merger times, more so than lower-e DNS.  Highly eccentric DNS have accelerated orbital decay due to strong gravitational wave emission at periastron and due to their early mergers are not likely to be represented in the observed sample (as is the case for our Galaxy). The importance of such short-lived systems has also been pointed out in the past \citep{BK2001, GRBB2005}.

Current estimates of Galactic DNS merger rates (of interest to ground-based gravitational-wave interferometers like LIGO) are based on the DNS observed sample and do not account for this selection effect against high-eccentricity DNS with short merger times (see \cite{Kal2004} and references therein). Consequently these rate estimates could underestimate the true merger rate by possibly a significant factor. This issue has in fact pointed out in the past in a somewhat different context (see \cite{BK2001} and \cite{Ivanova}). To assess the magnitude of this upward correction factor that should in principle be applied to such rate estimates, we calculate the fraction of DNS with merger times (at birth) shorter than 1\,Myr and 10\,Myr. We find that in most models these fractions are $\simeq10-15\%$ and $\simeq20-30\%$, respectively; for a couple of our extreme models (C and E1) the fractions are $\simeq50$\% for a cut of 10\,Myr. We conclude that although orbital evolution due to gravitational radiation affects the eccentricity distribution of the observed sample, the associated upwards correction factor to merger rate estimates is rather small (typically 10-40\%). This factor is well within the current uncertainties of the DNS merger rate estimates for the Galaxy.  

\acknowledgments
We thank Bart Willems for useful discussions and the referee Matthew Bailes for suggestions that improved the paper. This work has been supported by NSF Gravitational Physics grant PHYS-0353111 and a David and Lucile Packard Foundation Fellowship in Science and Engineering to VK, and KBN Grant 1P03D02228 to KB.



\end{document}